\documentclass[manuscript]{aastex}

\usepackage[dvips]{color}

\shorttitle{Critical Core Masses for Gas Giants}
\shortauthors{Hori & Ikoma}

\begin{document}

\title{Critical Core Masses for Gas Giant Formation with Grain-Free Envelopes}

\author{Yasunori Hori and Masahiro Ikoma}
\affil{Department of Earth and Planetary Sciences, Tokyo Institute of Technology, 
Ookayama, Meguro-ku, Tokyo 152-8551, Japan}

\email{hori@geo.titech.ac.jp}

\begin{abstract}
We investigate the critical core mass and the envelope growth timescale,
assuming grain-free envelopes, to examine how small cores are allowed to form gas giants
in the framework of the core accretion model.
This is motivated by a theoretical dilemma concerning Jupiter formation:
Modelings of Jupiter's interior suggest that it contains a small core of $< 10M_\oplus$,
while many core accretion models of Jupiter formation require a large core of 
$> 10M_\oplus$ to finish its formation by the time of disk dissipation.
Reduction of opacity in the accreting envelope is known to hasten gas giant formation.
Almost all the previous studies assumed grain-dominated opacity in the envelope.
Instead, we examine cases of grain-free envelopes in this study.
Our numerical simulations show that an isolated core of as small as $1.7 M_\oplus$
is able to capture disk gas to form a gas giant on a timescale of million years,
if the accreting envelope is grain-free;
that value decreases to $0.75 M_\oplus$,
if the envelope is metal-free, namely, composed purely of hydrogen and helium.
It is also shown that alkali atoms, which are known to be one of 
the dominant opacity sources near $1500 \mathrm{K}$ in the atmospheres of hot Jupiters,
have little contribution to determine the critical core mass.
Our results confirm that sedimentation and coagulation of grains in the accreting envelope
is a key to resolve the dilemma about Jupiter formation.
\end{abstract}


\section{Introduction\label{sec: intro}}

Core masses provide clues to unveiling the origins of gas giants.
In the core accretion scenario for gas giant formation,
a forming solid protoplanet (i.e., a proto-core)
experiences a rapid gas capture from a protoplanetary disk to form a massive
gas envelope, when its mass exceeds a critical mass \citep{hm80,bp86}.
That critical core mass must be reached within the lifetime of the disk
gas of several million years \citep[e.g.][]{h01}, which places a limit
on the core mass of a formed gas giant.
Because of the slow increase in the critical core mass
with core accretion rate \citep{ds82,mi00},
faster formation in general results in larger core mass.
Indeed, in many core-accretion models that are successful in
forming Jupiter within several Myr \citep[][etc.]{jbp96,iiw03,alibert05},
the resultant core mass is as large as $\sim 10M_\oplus$ or more.

In contrast, the mass of Jupiter's present core is inferred to be small.
\citet{sg04} made an extensive investigation of the interior structure of Jupiter,
finding successful models that are consistent with the observed values of its
gravitational moments and equatorial radius by using a
variety of equations of state (EOSs) for hydrogen and helium.
They demonstrated that the possible core mass of Jupiter is smaller than
$\sim 10~M_\oplus$.
This is also supported by recent calculations with an {\it ab initio} EOS
derived in the first-principle approach \citep{nn08}.
While a more massive core of $>10 M_\oplus$ is reported by \citet{bm08}
who used their own EOS of hydrogen-helium
mixtures based on density functional molecular dynamics, \citet{fn09} pointed out
the difference in the mass fraction of helium used by the two groups
is responsible for this discrepancy in the  derived value of Jupiter's core mass.
Although this pending problem about Jupiter's core may arise from
the uncertainty of EOS, the core mass suggested by interior modeling is, on an average,
smaller than that derived by formation theories.
This fact motivates us to know how small
a core can start the rapid gas accretion to form a massive
envelope within several Myr.

Reduction of opacity in the protoplanet's envelope
has the potential to make a small core possible.
As the opacity becomes small, the critical core mass decreases
and the post-critical-mass gas accretion becomes fast,
because low opacity in the envelope makes it difficult to
maintain the envelope's hydrostatic structure
without gravitational energy released by contraction of the envelope
\citep{hm80,ds82,mi00}.
Since the opacity sources are dust grains and gaseous components in
the envelope, a minimum critical core mass is achieved
in the case of grain-free envelopes.
All the previous studies except one
calculation done by \citet{hm80} (see Section~\ref{sec: gas opacity})
assumed grain-dominated opacity in the outer envelope.
Thus, in this paper, we consider grain-free envelopes and
make an extensive investigation of the critical core mass
and timescale for gas accretion.

Results from the recent works by Podolak and his colleagues
\citep[][]{mp03,mp08} are encouraging.
They have directly simulated the dynamical behavior of dust grains to determine
their size distribution, and then calculated grain opacity in the accreting envelope.
Their numerical simulations revealed that grain opacities in the envelope
can be much lower than those in the protoplanetary disk.
The reason is that small grains initially
suspended in the outer envelope quickly grow large in size and
then settle down into the deep envelope where temperature
is high enough that grains evaporate.

In the following section, the details of the gas opacity used in this study are described.
Our numerical calculations and results for the critical core mass and
the timescale of gas accretion are shown in Sections 3 and 4, respectively.
We discuss a possibility of Jupiter formation with a small core in Section~5.

\section{Gas Opacity \label{sec: gas opacity}}
The opacity of the envelope gas is lowest when the envelope
contains only hydrogen and helium.
In this study, we first consider such a \textit{metal-free} 
case\footnote{In the community of astrophysics, the word
`metal' represents elements heavier than helium.}. 
We compute chemical equilibrium between $\mathrm{H}_2$, H, $\mathrm{H}^+$, 
$\mathrm{H}^-$, $\mathrm{H}^+_2$, $\mathrm{H}^+_3$, and $\mathrm{e}^-$, 
and then calculate the opacity that includes the bound-free
and free-free absorptions by $\mathrm{H}^-$,
Rayleigh scatterings by $\mathrm{H}_2$, H, and He,
Thomson scattering by $\mathrm{e}^-$, and the collision-induced
absorptions (CIA) due to $\mathrm{H}_2\mathrm{-H}_2$,
$\mathrm{H}_2\mathrm{-H}$, $\mathrm{H}_2\mathrm{-He}$, and
$\mathrm{H}\mathrm{-He}$.
Values of quantities relevant to those calculations are given in
\citet{pl91} and references therein.
The CIA opacities are computed by using the latest programs and tables available
on Borysow's web
page\footnote{http://www.astro.ku.dk/~aborysow/programs/index.html}
\citep{ab85,ab90,b91,b92,zb95,gb96,ab97,b00,b02}.

Metal-containing envelopes with no grains are also considered in this study.
It is uncertain what fraction of metals in protoplanetary disks are
incorporated in dust grains and what faction remains in their gaseous forms.
For example, the amount of water adsorbed onto grain surfaces depends strongly on
the thermal state of the protoplanetary disk.
Water is adsorbed onto grain surfaces in a relatively cold disk,
while it remains in the gas phase in a relatively hot disk \citep{ajm02,ww09}.
Adsorption rates of water are also sensitive to the abundances and sizes of
dust grains in the disk \citep{an06}, but such dust properties remain poorly known.
Alkali atoms may be important:
They are known to be one of the dominant gas opacity sources
near $1500$K in the outer atmospheres
of brown dwarfs and giant planets \citep{tg94,ab00},
though alkali atoms have a tendency to reside onto grain surfaces
in protoplanetary disks \citep{hh93}.
Therefore, we consider two cases in addition to the metal-free case:
In one case called the alkali case hereafter,
alkali atoms are all in the gas phase;
in another case called the no-alkali case,
they are absent in the gas.

In the alkali case, we use the gas-opacity data provided by \citet{rsf08}, which 
incorporate the revised solar abundances and alkali atoms. 
The solar abundances have been recently revised
by re-analyses of two forbidden lines of neutral oxygen and carbon
from the solar photosphere under the assumption of the local
thermodynamic equilibrium, although their values should still require
scrutiny because of uncertainties in 3-D hydrodynamical models for the
solar atmosphere \citep{ap01,ap02,kl03}.
Opacity data without alkali atoms are kindly provided by Dr. Freedman
(in personal communication).
In practice, we use the up-to-date opacity data of \citet{rsf08}, which
cover the wider ranges of temperature and pressure
compared to the published ones
and include the latest HITRAN spectroscopic data and an
additional opacity source, $\mathrm{CO}_2$ (in personal communication).

\section{Critical Core Mass\label{sec: Mcrit}}

The spherically-symmetric hydrostatic structure of a protoplanet is simulated
in a manner similar to previous studies \citep[e.g.][]{hm80}.
The protoplanet consists of a solid core and a gaseous envelope.
The core has a constant density of $3.2 \mathrm{g}~\mathrm{cm}^{-3}$
and its structure is not computed. The envelope is assumed to be in purely
hydrostatic equilibrium and have a uniform chemical composition.
While the composition depends on the case, the abundance ratios of elements
taken into account are always solar \citep{kl03}.
All simulations are computed with spline-interpolated SCVH EOS tables
for hydrogen and helium provided by \cite{ds95}
, contributions from heavy elements being ignored.
The inner boundary conditions are applied at the core surface.
At the outer boundary,
the temperature and density are equal to the midplane values of a protoplanetary disk
at the protoplanet's orbit, $T_\mathrm{disk}$ and $\rho_\mathrm{disk}$,
because the envelope is assumed to be in equilibrium with the disk gas
at the outer edge.
The outer radius of the protoplanet is defined by the smaller of the accretion radius and
the tidal radius. Note that our results are insensitive to the outer boundary conditions.
We handle the accretion rate of planetesimals, $\dot{M}_\mathrm{c}$,
as a free parameter in this study.
We follow the same procedure as \citet{hm80} to determine the critical core mass, 
$M_\mathrm{crit}$ (please see \citet{hm80} in detail).
Table 1 summarizes input parameters and their values used in this study.

Figure 1 shows $M_\mathrm{crit}$ as a function of $\dot{M}_\mathrm{c}$ for the three cases.
For comparison, we plot results for additional two cases
in which the envelope contains both grains and gas as opacity sources. 
We assume that the grain opacity takes $f$ times values of the grain opacity 
given by \citet{jbp85};
their calculations assumed a nearly interstellar size distribution of dust grains.

Reduced opacity results in small $M_\mathrm{crit}$, as shown in Figure 1.
For instance, we now focus on results for 
$\dot{M}_\mathrm{c} = 1\times10^{-6}M_\oplus\mathrm{/yr}$.
In the case of $f = 0.01$ (double dot-dashed line),
which is often adopted in core accretion models \citep[e.g.][]{oh05},
$M_\mathrm{crit} = 10M_\oplus$.
Removal of grains lowers$M_\mathrm{crit}$ to
a few $M_\oplus$; $3.5M_\oplus$ for the alkali case (dashed line) and 
$1.5M_\oplus$ for the metal-free case (solid line).

Compared to the alkali case,
the metal-free case always produces smaller $M_\mathrm{crit}$.
This is because molecules composed of oxygen and/or carbon 
such as $\mathrm{H}_2\mathrm{O}$ and $\mathrm{CO}_2$ are effective opacity sources 
as for determination of $M_\mathrm{crit}$.
In contrast, comparison between the results for the alkali
(dashed line) and no-alkali cases (dotted line)
demonstrates that alkali atoms have little contribution to determine $M_\mathrm{crit}$.
The reason is that convection governs heat transfer 
in deep, hot parts of the envelope
where alkali atoms have a great contribution to the opacity.

Figure 1 also demonstrates that all the lines except one for the metal-free case are
converging, as $\dot{M}_\mathrm{c}$ decreases.
In the case of low $\dot{M}_\mathrm{c}$, namely, low luminosity,
the outermost isothermal layer extends deep in the envelope.
In that isothermal layer, the density increases rapidly to keep the
pressure gradient needed for supporting the core's gravity.
Thus, because of high densities, the opacity from
$\mathrm{H}_2 \mathrm{O}$ and $\mathrm{CO}_2$ dominates the grain opacity
in the deep radiative envelope in spite of low temperature.

Finally it is worth mentioning differences from previous studies.
\citet{hm80} calculated $M_\mathrm{crit}$ in 
the metal-free case for only one value of core accretion rate, 
$1\times10^{-6} M_\oplus \mathrm{/yr}$, and derived $1.5 M_\oplus$.
While the CIA opacities at high temperatures
have been revised thanks to progresses of quantum mechanical models and 
improvement of experimental data in the 1990s, Mizuno's (1980) value 
of $M_\mathrm{crit}$ is in good agreement with our result for the same parameters.
\citet{mi00} investigated cases with low grain opacity.
Our calculations for $f = 0.01$ and $f = 0.001$ yield larger $M_\mathrm{crit}$
than those from \citet{mi00},
because they assumed relatively small grain opacities compared to those used in this study.

\section{Envelope Growth Timescale\label{sec: tgas}}

The critical core mass decreases as the core accretion rate decreases, as shown in Fig.1.
Therefore, the case of $\dot{M}_\mathrm{c} \rightarrow 0$
(i.e., isothermal envelope) yields the absolute minimum of $M_\mathrm{crit}$
\citep{sasaki89,pecnik05}.
However, for smaller $M_\mathrm{crit}$, it takes longer for the core to
capture disk gas \citep{mi00,ikoma_genda06}.
Because there are time constraints on gas giant formation such as
the disk's lifetime \citep{h01}, a practical minimum of $M_\mathrm{crit}$
is determined in this respect.
It is necessary to estimate how long it takes for a protoplanet with a given core mass, 
$M_\mathrm{core}$, to capture disk gas.

We consider an isolated protoplanet, namely,
accumulation of the envelope after planetesimal accretion is halted.
The isolated protoplanet always experiences contraction of its envelope and 
captures disk gas, because of no energy supply due to planetesimal accretion.
We simulate the quasi-static evolution of the envelope
with a given $M_\mathrm{core}$ after planetesimal accretion is halted
and evaluate the growth timescale of the envelope.
As the growth timescale of the envelope,
we present values of the characteristic growth time, $\tau_\mathrm{g}$, 
that is defined by \citet{mi00}. 

Figure 2 plots $\tau_\mathrm{g}$ as functions of $M_\mathrm{core}$.
For comparison, the results for $f = 0.01$ and $0.001$ are also shown.
The growth time $\tau_\mathrm{g}$ is found to be much shorter in the case of
the grain-free opacities than $\tau_\mathrm{g}$ for the grain-dominated opacities.
For example, when $M_\mathrm{c} = 3M_\oplus$,
$\tau_\mathrm{g} = 7 \times 10^2~\mathrm{yr}$ in the metal-free case (solid line) and
$7 \times 10^4~\mathrm{yr}$ in the alkali case (dashed line),
while $2 \times 10^6~\mathrm{yr}$ for $f = 0.01$ (double dot-dashed line).
The envelope growth is regulated by the Kelvin-Helmholtz contraction of the envelope;
that is, $\tau_\mathrm{g}$ is proportional to $1/L$ \citep{mi00},
where $L$ is the luminosity at the envelope's outer edge.
Since a change in opacity compensates that in luminosity,
$\bar{\kappa}$ being an averaged opacity,
$\tau_\mathrm{g}$ should be proportional to $\bar{\kappa}/L$.
In the case of the grain-dominated opacities, the dependence is simple,
namely, $\tau_\mathrm{g} \propto f$, as shown in \citet{mi00} and
\citet{ikoma_genda06}.
Figure 2, however, demonstrates that such a simple scaling is inappropriate in low
$\bar{\kappa}$ cases.
The curves for grain-free cases are steeper than those for grain-dominated cases;
the reason has been already described in the previous section.

In any case, we have found that the envelope growth time is significant shorter
in the case of the grain-free opacities.
Based on this fact, we discuss the minimum critical core mass from the
viewpoint of gas giant formation, and the possibility of Jupiter formation with
a small core in the following section.

\section{Discussion\label{sec: discussion}}

There is a theoretical dilemma concerning Jupiter formation,
as described in Introduction.
Modelings of Jupiter's interior suggest that 
Jupiter has a small core of $<$~$10 M_\oplus$ \citep[e.g.,][]{sg04},   
while many core-accretion models of Jupiter formation require a large core
of $>$~$10 M_\oplus$
to finish its formation by the time of disk dissipation
\citep[][]{jbp96,alibert04,alibert05,fortier07,fortier09}.
The disk instability scenario has been revisited as
an alternative scenario of their formation \citep[e.g.][]{boss00,mayer02}.

In this study, we have demonstrated that
reduced opacities in the protoplanet's envelope
have the potential to resolve this dilemma.
From Fig.~\ref{fig2}, one finds that $M_{\rm core} = 0.75 M_\oplus$ 
for $\tau_g =$ 1~Myr in the metal-free case; $M_{\rm core} = 1.7 M_\oplus$ 
even in the alkali case.
Given observed lifetimes of protoplanetary disks of several Myr, 
the fact above indicates the reduction of opacity
allows Jupiter to have a small core that is consistent with interior modelings in principle.
The feasibility of such minimum $M_\mathrm{crit}$ depends on opacities
in the protoplanet's envelope, while this does not change our conclusion that
the minimum $M_\mathrm{crit}$ obtained here
provides the lowest limit to
core masses of gas giants to which the core accretion model can apply.
We need more extensive investigation of gas giant formation
into which sedimentation and coagulation of grains in the accreting envelope
are incorporated, although reduction of opacity was already pronounced \citep{mp03,mp08}. 

Our results also shed light upon growth of solid cores.
Many core accretion models assumed the presence of a single protoplanet
\citep[e.g.][]{jbp96}.
However, a gas giant is in practice thought of as being formed
in a system of multiple protoplanets embedded in a protoplanetary disk.
Compared to cases of a single protoplanet,
the final mass of a core should be small in the case of a multiple-protoplanet system.
According to \citet{ki98,ki00},
the isolation mass is a few $M_\oplus$ around $5$AU.
Even such a small core is enough for a gas giant
to capture disk gas within several Myr, as demonstrated in this study.
We need to perform comprehensive simulations on gas giant formation
in multiple-protoplanet systems, which incorporate adequate models of planetary accretion
such as fragmentation of planetesimals and $e$-damping due to gas drag
\citep[e.g.][]{duncan09}.
These calculations will be our future work.

\acknowledgments

We are grateful to S. Ida for his continuous encouragement.
We thank R.S. Freedman for his kindness of 
calculating newly Rosseland mean opacities of gas
with no alkali atom and providing updated gas opacities.
We also thank H. Tanaka and H. Nomura for giving us helpful comments.
Y.H. is supported by Grant-in-Aid for JSPS Fellows (No.21009495) from 
the Ministry of Education, Culture, Sports, Science and Technology (MEXT) of Japan.

\clearpage

\begin{figure}
\epsscale{.80}
\plotone{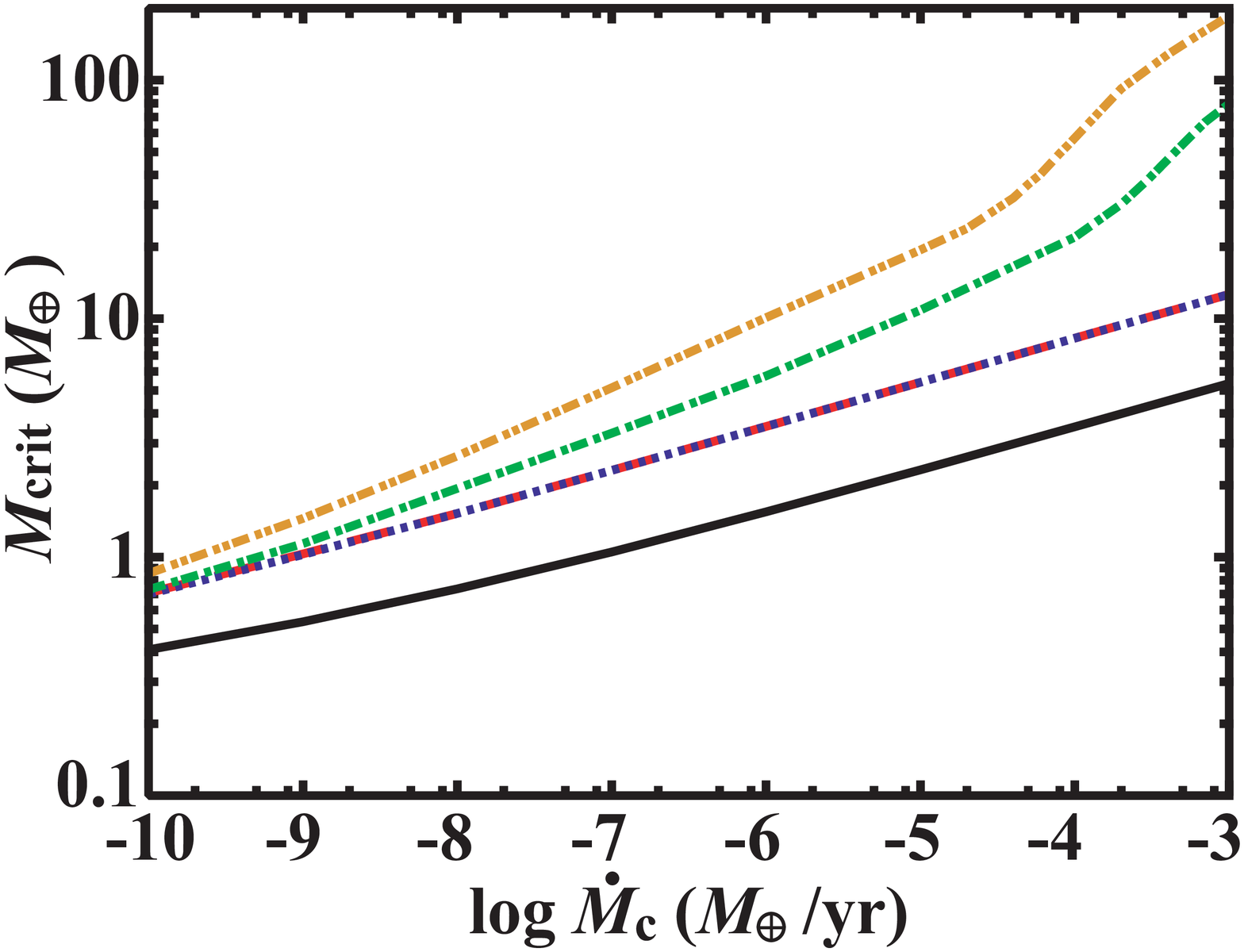}
\caption{Critical core masses, $M_\mathrm{crit}$, as functions of core accretion rate,
$\dot{M}_\mathrm{c}$.
The solid, dashed, and dotted lines 
represent the results for the metal-free, alkali, and no-alkali cases
(see Section \ref{sec: gas opacity} for the definitions).
For comparison, the results of two cases in which there exist grains in the envelope
are also shown with the double dot-dashed and dot-dashed lines:
$f = 0.01$ and $0.001$, respectively, where $f$ is the grain depletion factor
(see the text for its definition).
We have used the grain opacity tables derived from \citet{jbp85}.
[A color version of this plot is available
in the electronic edition of {\it The Astrophysical Journal}.\label{fig1}]}
\end{figure}

\clearpage

\begin{figure}
\epsscale{.80}
\plotone{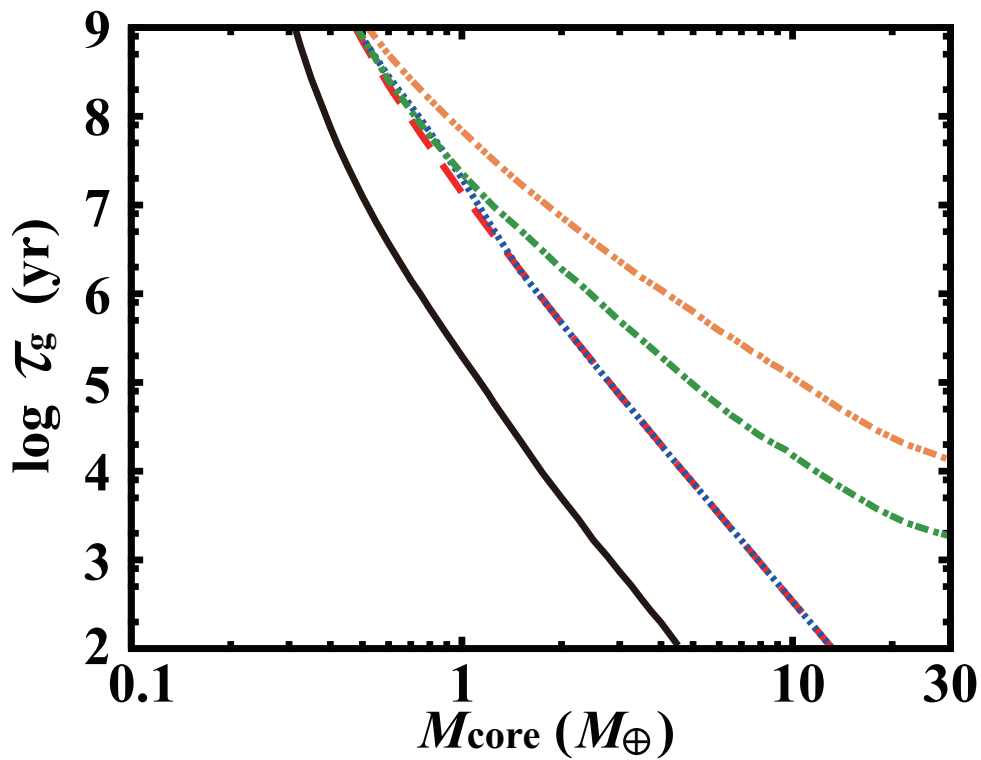}
\caption{Typical growth timescale of the envelope, $\tau_\mathrm{g}$
(see the text for the definition), as a function of $M_\mathrm{core}$.
As in Fig.1, the solid, dashed, dotted lines represent the results of
the metal-free, alkali, and no-alkali cases respectively.
The results of $f = 0.01$ (double dot-dashed line) and $0.001$ (dot-dashed line)
are also shown.
[A color version of this plot is available
in the electronic edition of {\it The Astrophysical Journal}.\label{fig2}]}
\end{figure}

\clearpage

\begin{table}
\caption{Input Parameters and Their Values.\label{tbl-1}}
\begin{flushleft}
\begin{tabular}{lrl}
\tableline\tableline
Parameter & &  Value \\
\tableline
semimajor axis, $a$ & & 5.2 AU \\
core density & & $3.2~\mathrm{g/cm^3}$ \\
disk temperature, $T_\mathrm{disk}$ & & $150~\mathrm{K}$ \\
disk density, $\rho_\mathrm{disk}$ & & $5.0\times 10^{-11}~\mathrm{g/cm^3}$ \\
planetesimal accretion rate, $\dot{M}_\mathrm{c}$ & & $1.0\times10^{-3}$~to~
$1.0\times10^{-10}$ 
$M_\oplus / \mathrm{yr}$ \\
\tableline
\end{tabular}
\end{flushleft}
\end{table}

\end{document}